# Proposing a Semantic Movie Recommendation System Enhanced by ChatGPT's NLP Results

Ali Fallahi Rahmatabadi, Azam Bastanfard *, Amineh Amini, Hadi Saboohi
Department of Computer Engineering
Karaj Branch, Islamic Azad University
Karaj, Iran
ali.fallahi@kiau.ac.ir, bastanfard@kiau.ac.ir, aamini@kiau.ac.ir, saboohi@kiau.ac.ir

*Abstract*— The importance of recommender systems on the web has grown, especially in the movie industry, with a vast selection of options to watch. To assist users in traversing available items and finding relevant results, recommender systems analyze operational data and investigate users' tastes and habits. Providing highly individualized suggestions can boost user engagement and satisfaction, which is one of the fundamental goals of the movie industry, significantly in online platforms. According to recent studies and research, using knowledge-based techniques and considering the semantic ideas of the textual data is a suitable way to get more appropriate results. This study provides a new method for building a knowledge graph based on semantic information. It uses the ChatGPT, as a large language model, to assess the brief descriptions of movies and extract their tone of voice. Results indicated that using the proposed method may significantly enhance accuracy rather than employing the explicit genres supplied by the publishers.

## I. INTRODUCTION

Our lives have experienced numerous fundamental changes in various areas due to the exponential growth of web-based technologies over the last decades. The digital era has multiple benefits and opportunities, such as E-learning [1], social media [2], online shops [3], messengers applications [4], etc. However, the continuously growing volume of data makes it a lot more complicated for users to access what they're looking for [5]. Recommendation systems consider the mentioned issue and provide customized suggestions to help users get the desired data [6].

Nowadays, recommender systems are used by users in numerous domains, including healthcare [7], tourism [8], business [9], movies [10], music [11], social network [12] etc. While various studies have used different methodologies to develop a recommendation system, the most often utilized techniques include content-based, context-aware, collaborative filtering, and hybrid [13; 14].

### A. Content-based

In this category takes advantage of the content-based characteristics of prior user selections to suggest related items in the dataset. The most common resources for creating a content-based recommender are textual resources [15].

### B. Context-aware

In a context-aware recommender system, contextual and demographic information, such as the user's location, gender, age and occupation, serves as the primary data source. These recommenders organize users based on shared characteristics and attributes and then provide the same recommendations to each group's members [16].

### C. Collaborative filtering (CF)

The fundamental idea of collaborative filtering (CF) is that future behavior would be comparable to those with similar interests and quantifiable activities such as comments, likes, and dislikes in the past [17]. Collaborative filtering is regarded as the most frequently prevalent technique for developing a recommender system [18].

### D. Hybrid recommenders

Creating a successful recommender system requires improving accuracy and offering more customized choices. Some studies have sought to build hybrid systems that combine various techniques and pattern formation methods [19].

Making an efficient recommender requires improving accuracy and providing more customized recommendations. However, sparsity and cold start are two of the most significant challenges in building efficient recommendation systems [20]. Data sparsity results from a lack of user interaction with the system [21]. Cold start is considered a situation when a new user or item has recently been added to the systems, and the recommender engine does not have any background information about the target [22].

Recent developments in deep learning and neural networks have opened up new opportunities for implementing systems based on approaches and concepts such as reinforcement learning [23] and transfer learning [24] to create recommenders that can handle enormous amounts of data on complex networks [25]. A large language model called ChatGPT [26] developed by OpenAI, which is built on the Generative Pre-trained Transformer (GPT) architecture, currently includes a number of significant online use cases.

The company released the GPT-1 in 2018 and the current version in 2023 is GPT-4. The capacity of ChatGPT to provide high-level replies to natural language queries is one of the main advantages of this model. The feature may enhance the user experience for a variety of online applications [27].

This work aims to propose a new method of using ChatGPT's natural language processing results to make a knowledge graph and build a semantic movie recommender system. Moreover, graphical visualization of the used dataset proposed to clarify some notable aspects of the users' tendencies in this domain. The remaining sections are grouped as follows: The second section looks at related literature. Section 3 provides the proposed method. The findings of the experiments are discussed in Section 4. Section 5 covers the specifics of our domain research findings, as well as any unresolved problems and future prospects. This section wraps up and summarizes our research.

state-of-the-art natural language processing module for building new categorizing. Moreover, provided graph knowledge and its visualization reveal significant characteristics about users. The following is a summary of the primary motivations for this study:

1) Introducing a new method for building a knowledge graph based on the tone of voice extracted from each movie's description's semantic information.

2) Providing a new method for building a knowledge graph based on the tone of voice extracted from each movie's description's semantic information.

3) Presenting a graphical visualization of the used dataset to clarify notable user tendencies in the movie domain.

## II. PREVIOUS WORK

In the movie industry, the genre is a critical key for producers and audiences [28]. Allocating time and financial resources to make a movie in a genre that is not popular is too risky for film companies. On the other hand, staying limited only to predefined genres may damage the audience's satisfaction. In this situation, categorizing movies based on a sentimental approach and providing personalized suggestions by considering each user's past activities and tendencies can benefit all film industry stakeholders.

In recent years, studies such as Delimayanti et al [29], Ranjith et [30], Nair and Cheriyan [31], and Reddy et al [32] have been done to develop movie recommendations based on genre data. Contrary to the present work, no research from previous years directly addressed sentiment analysis on movie descriptions and cutting-edge techniques like ChatGPT. Also, providing a comprehensive knowledge graph on the chosen dataset gives valuable information about the selected domain. In order to present a thorough examination, we also analyzed the previously related research and discussed their notable findings in the following lines.

The work presented in [29], proposed a content-based movie recommender based on features including title, genre, ratings, hashtags, and likes. The authors used the Term Frequency and Inverse Document Frequency (TF-IDF), K Nearest Neighbor (KNN) classification for analyzing the data, and the cosine similarity for finding matching users. Different approaches for finding similar users were utilized in [30]: user rating patterns, movie properties like genre, cast, and crew, and demographic data such as location, age, etc. Sentiment analysis techniques were used to determine the emotions concealed inside reviews. Moreover, a VGG-16 architecture [33] was employed to extract visual information from posters as another information resource. Likewise, explicit genre correlation in the combination of other elements, such as the user's favorite directors and ratings, to make a knowledge graph was operated in [31; 32].

While there are a few similarities in covering the concepts of movie genre between the mentioned publications and the current work, as a contribution, we utilized the ChatGPT as a

## III. PROPOSED METHOD

In the multimedia industry, recommender systems often use elements from structured sources, such as cast and genre, as well as less structured, including reviews and comments [34]. Using these characteristics presents two fundamental difficulties: first, being constrained to the stated words, and second, the sparsity issue, or the demand for additional data resulting from insufficient user contact with the system. Extracting semantic and implicit characteristics like the tone of voice can enhance the recommendations and user satisfaction. We provide a step-by-step explanation of the proposed method in the following paragraphs. The main architecture, including different parts and their relations are shown in *Fig 1*.

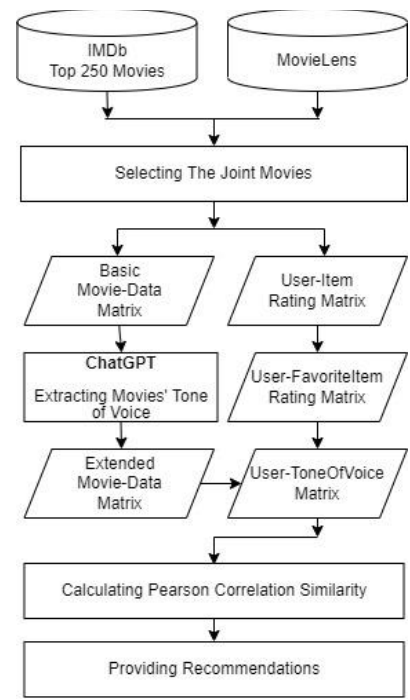

Fig. 1. The main architecture of the proposed method.

Tow datasets were used as the primary inputs of the proposed recommender. The first one contains 250 movies from the IMdB website (https://www.imdb.com/). These movies are presented as the IMDb Top 250 as rated by regular IMDb voters. The second dataset, Movielense [35], was introduced by the University of Minnesota's GroupLens research project. Movies descriptions and genres were crawled from IMDb. Ratings for the selected movies were collected from Movielense. Ratings ranged from 1 to 5, with 5 being the greatest enjoyment. Due to the incompatibility between the two datasets, 241 movies could be chosen as the input data.

Two matrices were generated from the input data. The Basic Movie-Data matrix has movie ids from Movielense as row values and genres from IMDb. Cells can be 1, meaning the target movie has that genre., or 0, meaning opposite.

The second matrix, User-Item rating, has user ids in its rows and movie ids in its columns. Cells in this matrix are rates from users with values from 0 to 5. In this matrix all the values are form Movielense.

In the next step, movie descriptions from the Basic Movie-Data matrix are sent to the ChatGPT for analyzing their tone of voice. The results were added to the Basic matrix. So, the Extended Movie-Data matrix was produced.

Values on the User-Item rating matrix were also filtered only to retain the rates greater than three from the maximum rating value, which is 5. The result of this process is the User-FavoriteItem matrix.

After that, the tone of voice values from the Extended matrix were combined with the User-FavoriteItem matrix to produce the User-ToneOfVoice matrix. The result matrix has 6938 rows (users), 127 columns (tone of voice), and 180,040 active cells(rates).

The Pearson Correlation Coefficient (PCC)formula was run on the User-FavoriteItem matrix values to calculate similarities. The PCC formula is shown in the formula (1).

$$PCC_{a,b} = \frac{\sum_{i=1}^{I_{a,b}}(r_{a,i} - \overline{r}_a)(r_{b,i} - \overline{r}_b)}{\sqrt{\sum_{i=1}^{I_a}(r_{a,i} - \overline{r}_a)^2}\sqrt{\sum_{i=1}^{I_b}(r_{b,i} - \overline{r}_b)^2}} \quad (1)$$

In the PCC formula, $r_{a,i}$ denotes the rating score from the target user $a$ for item $i$, and $r_{b,i}$ indicates the rating score from the target user $b$ for item $i$. Also, $\overline{r}_a$ and $\overline{r}_b$ provide the average ratings of users $a$ and $b$ based on all the objects each user has rated.

Finally, suggestions will be generated based on the collaborative filtering approach and considering the top N users who are the most similar to the target user.

## IV. EXPRIMENTAL RESULT

In this section, the outcomes of the proposed method are showcased. Moreover, a detailed dataset investigation is conducted by analyzing relevant perspectives. Experiments were conducted in five different phases based on the input data.

### A. Evaluation metrics

Mean Absolute Error (MAE) and Root Mean Square Error (RMSE) were employed as two evaluation metrics. The MAE calculates the difference between the original user rating and the calculated rating by the recommender. Adds all these differences and divides the result by the number of items taken into consideration [36]. The MAE measure is shown in Formula 2.

$$\text{MAE} = \frac{1}{n}\sum_{i=1}^{n}|P_{(u,i)} - R_{(u,i)}| \quad (2)$$

In equation (2), $P_{(u,i)}$ stands for the predicted rating for the item $i$. from target user $u$. $R_{(u,i)}$ stands for the first score. The total number of ratings for the item set is $N$.

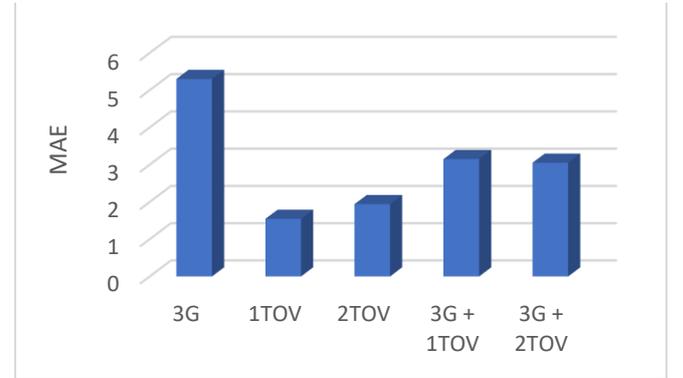

Fig. 2. Diversity of the movies based on their tone of voices.

Fig 2 presents the proposed method evaluation by the MAE metric in five different setups. In fig 2:

- 3G stands for the MAE result when the input of the recommender system included only genres provided by IMDb. On the IMDb website, each movie has one to three genres.

- 1TOV and 2TOV indicate when the input was one or two tones of voice.

- 3G+1TOV represented when the input data included three genres from the IMDb and one tone of voice. Likewise, 3G+2TOV defines three genres from the IMDb and two tones of voice.

Among the five mentioned situations, the best results were produced when the input data included only one tone of voice for each movie. This setup is labeled with 1TOV in figure2.

The RMSE highlights absolute inaccuracy. Therefore when the RMSE is low, the recommendation accuracy is considered adequate [37]. The RMSE measure is shown in Formula 3.

$$\text{RMSE} = \sqrt{\frac{\sum_{i=1}^{n}(P_{(u,i)} - R_{(u,i)})^2}{N}} \quad (3)$$

In equation (3), $P_{(u,i)}$ is the predicted rating for item $i$ from target user $u$. The original score is denoted by $R_{(u,i)}$. The item set has $N$ total ratings in total.

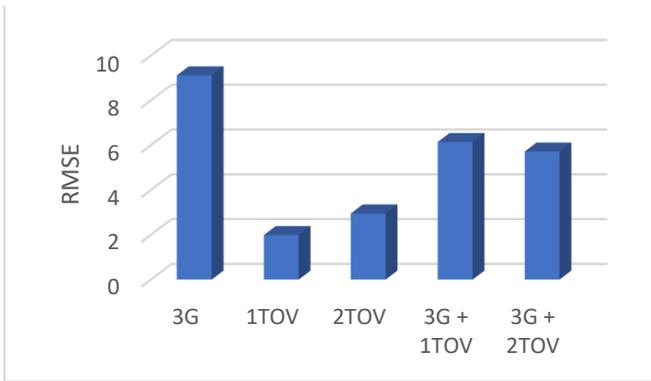

Fig. 3. Diversity of the movies based on their tone of voices.

Fig. 3 shows the RMSE metric evaluation of the proposed method in five configurations. In fig. 3, 3G is the MAE result when the recommender used solely IMDb genres. Each IMDb movie has one to three genres. 1TOV and 2TOV signify one or two terms as tone of voice for the input. 3G+1TOV represented input data with three IMDb genres and one tone of voice. 3G+2TOV specifies three IMDb genres and two voice tones.

The most excellent results were obtained when each movie had just one tone of voice. Figure 2 labels this arrangement 1TOV. To clarify the provided results in fig 2 and fig 3, precise values of the proposed method's evaluation based on MAE and RMSE are presented in *Table 1*.

TABLE I. EVALUATION OF THE PROPOSED METHOD BASED ON MAE AND RMSE.

| Evaluation Metric<br>Input Data | MAE | RMSE |
| --- | --- | --- |
| 3 Genres from IMDb (21 item) | 5.3011 | 9.1198 |
| Tone of Voice (80 item) | 1.5476 | 1.9896 |
| Tone of Voice (126 item) | 1.9757 | 2.9542 |
| 3 Genres + 80 Tone of Voice | 3.1507 | 6.1451 |
| 3 Genres + 126 Tone of Voice | 3.0513 | 5.7193 |

Based on the presented results in table 1, it can be concluded that using extracted tone of voice from movie descriptions is much more suitable than only categorizing movies based on predefined genres for building a movie recommender system.

### B. Dataset overview

The chart *Fig 4*, is produced based on 21 unique genres from IMdB that were used 720 times for the movies in the dataset. These genres are: Action, Adventure, Animation, Biography, Comedy, Crime, Drama, Family, Fantasy, Film Noir, History, Horror, Music, Musical, Mystery, Romance, Sci-Fi, Sport, Thriller, War, Western.

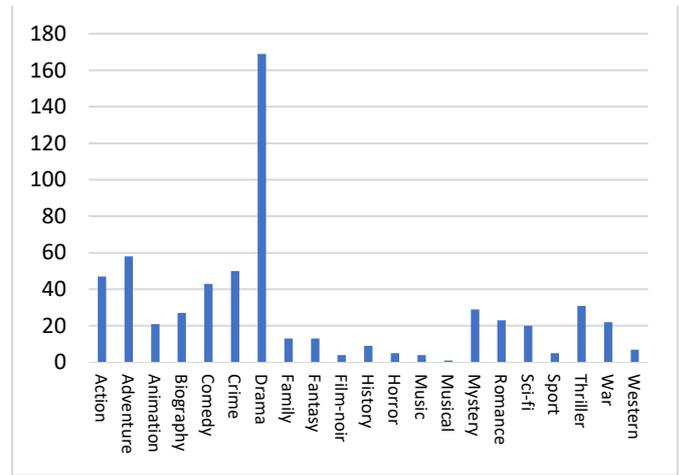

Fig. 4. Diversity of the movies based on the IMdB genres.

As presented in *Fig 4*, The top three genres are Drama (169 items), Adventure (58 items), and Action (47 items). The last genres, respectively, are: Musical (1 item), Music (4 items), and Film-noir (4 items).

The pie chart in *Fig 5* shows the result of the analysis of the movies based on their tone of voice. Totally the ChatGPT produced 126 unique terms as tone of voice for 241 movies. To provide a more practical presentation, tone of voices with the same frequency and less than five were combined into one group, so the chart has twenty-two divisions.

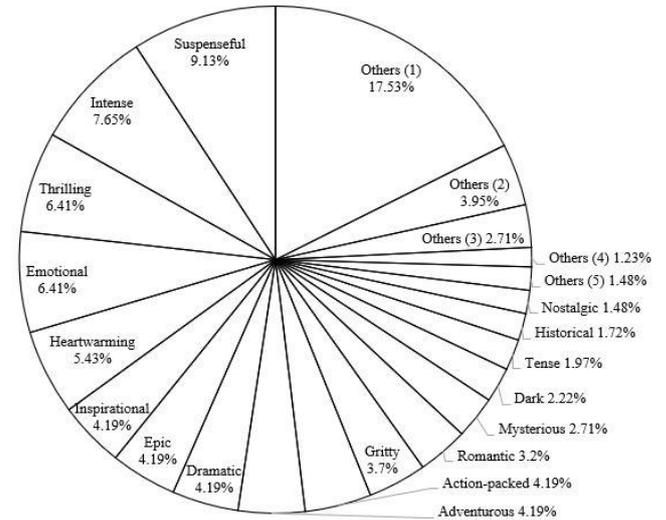

Fig. 5. Diversity of the movies based on their tone of voices.

According to the findings in the dataset, the most favorite movie's tone of voice is Suspenseful. After that, Intense is the second favorite tone of voice among the most liked movies. Respectively Emotional and Thrilling by the quantity of 26 put in third place in this comparison.

To provide another perspective on the values listed in *Fig 5* the quantities of the tone of voices are presented in detail in *Table 2*.

TABLE II. ANALYSING THE MOVIES BASED ON THEIR TONE OF VOICE

| Term | Quantity | Percentage |
| --- | --- | --- |
| Others (1) | 71 | 17.53 % |
| Suspenseful | 37 | 9.13 % |
| Intense | 31 | 7.65 % |
| Emotional | 26 | 6.41 % |
| Thrilling | 26 | 6.41 % |
| Heartwarming | 22 | 5.43 % |
| Action-packed | 17 | 4.19 % |
| Adventurous | 17 | 4.19 % |
| Dramatic | 17 | 4.19 % |
| Epic | 17 | 4.19 % |
| Inspirational | 17 | 4.19 % |
| Others (2) | 16 | 3.95 % |
| Gritty | 15 | 3.7 % |
| Romantic | 13 | 3.2 % |
| Others (3) | 11 | 2.71 % |
| Mysterious | 11 | 2.71 % |
| Dark | 9 | 2.22 % |
| Tense | 8 | 1.97 % |
| Historical | 7 | 1.72 % |
| Others (5) | 6 | 1.48 % |
| Nostalgic | 6 | 1.48 % |
| Others (4) | 5 | 1.23 % |

Although to make the presentation more practical, the tone of voices with a similar frequency and a count of less than five were grouped in *Fig 5* and *Table 2*. However, detailed information is provided in the following lines for possible additional studies.

- Others (1): Contemplative, Superheroic, Compassionate, Silly, Shocking, Complex, Grieving, Creepy, Sad, Crime, Road-trip, Riveting, Cynical, Repetitive, Surprising, Swashbuckling, Comical, Unconventional, Ambitious, War-drama, Betrayed, Vengeful, Uplifting, Unusual, Underdog-story, Bittersweet, Dangerous, Touching, Brave, Threatened, Cat-and-mouse, Chaotic, Charming, Claustrophobic, Regretful, Darkly Comedic, Redemptive, Horror, Investigative, Dystopian, Eerie, Empowering, Envious, Idealistic, Exhilarating, Legal, Family-oriented, Heroic, Fantasy, Farce, Festive, Harrowing, Jealous, Lonely, Rebellious, Patriotic, Darkly humorous, Provocative, Deliberative, Poignant, Playful, Philosophical, Desperate, Mad, Noir, Determined, Disturbed, Motivating, Divergent, Magical, Intimate

- Others (2): Introspective, Moving, Thoughtful,

- Greedy, Realistic, Heart-wrenching, Heartfelt, Somber, Witty, Exciting, Disturbing, Futuristic, Deceptive, Fantastical, Coming-of-age, Enchanting

- Others (3): Terrifying, Hopeful, Mind-bending, Melancholic, Political, Revengeful, Satirical, Sci-fi, Serious, Competitive, Quirky

- Others (4): Violent, Tragic, Humorous, Psychological, Surreal

- Others (5): Comedic, Intriguing, Thought-provoking, Reflective, Whimsical, Heartbreaking

## V. CONCLUSION

The increasing expansion of online information has made it challenging for users to get the data they need. Generally, recommender systems are utilized in this context as valuable tools for information filtering, helping individuals make more beneficial decisions by providing tailored suggestions. However, the two biggest challenges to delivering proper recommendations are sparsity and cold start.

This work offers a new method for developing a semantic movie recommendation engine. Unlike other research, we extracted the tone of voice from the short descriptions of movies using ChatGPT as a large language model. Results indicated that using the proposed method may significantly enhance accuracy rather than employing the explicit genres provided by the publishers.

Investigating the domain area and the proposed method's achievements illustrate that implicit semantic data, like the movie's description's tone of voice, may be very effective at increasing the accuracy and enhancing the performance of recommendation systems. For future work, analyzing the titles, songs, and lyrics of the films may be other valuable resources to improve semantic movie recommendation systems.